\documentclass[amsmath,amssymb,preprintnumbers,nofootinbib,prd,a4paper,twocolumn]{revtex4-1}
\pdfoutput=1
\usepackage{amsthm}
\usepackage{amsmath}
\usepackage{graphicx}
\usepackage{bbold}
\usepackage{color}
\usepackage{subfigure}
\usepackage{multirow}
\usepackage{graphicx,bm}

\usepackage{dcolumn}
\usepackage{bm}
\usepackage{slashed}
\usepackage{epsfig}
\usepackage{hyperref}
\usepackage{verbatim}

\usepackage[utf8]{inputenc}
\usepackage[T1]{fontenc}
\usepackage[normalem]{ulem}

\newcommand{\ii}{\mathrm{i}}
\newcommand{\beq}{\begin{eqnarray}}
\newcommand{\eeq}{\end{eqnarray}}

\newcommand{\bmp}{\noindent\begin{minipage}{16cm}}
\newcommand{\emp}{\end{minipage}\vskip 7mm} 

\newcommand{\be}{\begin{eqnarray}}
\newcommand{\ee}{\end{eqnarray}}

\newcommand{\SU}{\mbox{SU}}
\newcommand{\SO}{\mbox{SO}}
\newcommand{\SP}{\mbox{Sp}}
\newcommand{\UU}{\mbox{U}}

\makeatletter
\DeclareTextCommand{\textprime}{\encodingdefault}{%
  \mbox{$\m@th'\kern-\scriptspace$}%
}
\makeatother

\begin{document}
\preprint{CP3-Origins-2020-02 DNRF90}
\preprint{FERMILAB-PUB-20-055-T}

\title{Natural Top-Bottom Mass Hierarchy in Composite Higgs Models}


\author{Martin {\sc Rosenlyst}}
\email{rosenlyst@cp3.sdu.dk}
\affiliation{CP$^3$-Origins, University of Southern Denmark, Campusvej 55, DK-5230 Odense M, Denmark}

\author{Christopher {\sc T. Hill}}
\email{hill@fnal.gov}
\affiliation{Fermi National Accelerator Laboratory, P.O. Box 500, Batavia, Illinois 60510, USA\\}

\date{\today}


\begin{abstract}
We consider composite two-Higgs doublet models based on 
gauge-Yukawa theories with strongly interacting fermions generating 
the top-bottom mass hierarchy. The model features a single ``universal'' Higgs-Yukawa 
coupling, $ g $, which  is identified with the top quark 
$ g\equiv g_t \sim \mathcal{O}(1)  $. The top-bottom mass 
hierarchy arises by soft breaking of a $ \mathbb{Z}_2 $ 
symmetry by a condensate of strongly interacting fermions. 
A mass splitting between vector-like masses of the confined techni-fermions 
controls this top-bottom mass hierarchy. 
This mechanism can be present in a variety of models based on vacuum misalignment. 
For concreteness, we demonstrate it in a composite two-Higgs scheme. 
\end{abstract}

\maketitle

\section{Introduction}

The relative hierarchy of the top mass to the other SM fermion masses 
has long been of interest~\cite{PR,FP,Yama,BHL,He:2001fz,Porto:2007ed,BenTov:2012cx,Hill:2014mqa,Hill:2019ldq,Hill:2019cce}.
The mass of the top quark is of the same order as the electroweak symmetry breaking 
(EWSB) scale ($ v_{\rm EW} = (\sqrt{2}G_F)^{-1/2} \simeq 246 $ GeV), while in some 
``leading order'' the
other fermion masses small. In the SM, the observed mass spectrum of the fermions is achieved 
by inputting a hierarchy of the Higgs--Yukawa (HY) coupling constants, and 
while technically natural, no dynamical rationale for the hierarchy is offered.

In this letter we explore a mechanism for generating the top-bottom mass 
hierarchy with a universal HY coupling constant ($ g\equiv g_t = g_b \sim \mathcal{O}(1) $).
Our background interest is multi-Higgs models, composites due to something like a new
strong force or gravity, which is flavor blind and generates a large spectrum 
with universal HY couplings.  The resulting fermion mass hierarchies and CKM matrix 
are then relegated the problem of the multi-Higgs masses and mixings~\cite{Hill:2019ldq,Hill:2019cce}.  
We explore a realization of this in a composite two-Higgs doublet models (2HDM) with 
soft breaking of an associated $ \mathbb{Z}_2 $, and based on a new strong dynamics.  
Here the Higgs bosons arise as pseudo-Nambu-Goldstone bosons (pNGB) from a spontaneously 
broken global symmetry. The geneology
of Higges-as-pNGB's includes  ``Composite Higgs'' (CH)~\cite{Kaplan:1983fs}, ``partially Composite Higgs'' 
(pCH)~\cite{Galloway:2016fuo,Alanne:2017rrs,Alanne:2017ymh},
``Little Higgs''~\cite{ArkaniHamed:2001nc,ArkaniHamed:2002qx}, 
``holographic extra dimensions''~\cite{Contino:2003ve,Hosotani:2005nz}, 
``Twin Higgs''~\cite{Chacko:2005pe} and 
``elementary Goldstone Higgs models''~\cite{Alanne:2014kea}.

We start with two composite Higgs doublets $ H_t $ and $ H_b $ that are, 
respectively, even and odd under a discrete $ \mathbb{Z}_2 $ symmetry. 
The Higgs doublet $ H_t $ is identified with the observed SM doublet, where 
its neutral, $ H_t^0 $, develops the electroweak (EW) vacuum expectation value (VEV). 
A small ``tadpole'' 
VEV can be induced in
$ H_b^0 $ by introducing a $ H_t $-$ H_b $ mixing term that softly breaks 
$ \mathbb{Z}_2 $ symmetry. 

By demanding the left- and right-handed 
top and bottom quarks transform as 
$ Q_{L,3} \equiv (t_L,b_L)^T \rightarrow Q_{L,3}$, $ t_R \rightarrow t_R $ and 
$ b_R\rightarrow -b_R $ under the $ \mathbb{Z}_2 $ symmetry, the $ H_t^0 $ 
only couples to the top quark while $ H_b^0 $ only to the bottom quark. 
The HY couplings are generated by effective 
operators involving the quarks and the 
composite Higgs doublets 
with one universal HY coupling constant, $ g\sim \mathcal{O}(1) $. 
The mass hierarchy $ m_t/ m_b \approx 40  $ is 
thus generated naturally without fine-tuning.  

For concreteness, we 
consider the minimal composite 2HDM fulfilling the above requirements, the 
$\SU(6)/\SP(6)$ model of~\cite{Cai:2018tet}. In this model, 
the top-bottom mass hierarchy can be controlled by a mass 
splitting in the explicit vector-like masses 
of the confining fermions. This mass splitting 
also leads to a heavy isodoublet $ H_b^0 $ ($ \sim\mathcal{O}(1) $ TeV), which is possibly observable at LHC or its upgrades. The model dynamically generates the top-bottom mass hierarchy 
and simultaneously alleviates the hierarchy problem of EWSB. 

Finally, we will briefly consider ideas to future studies such as: 
(i) making an UV complete theory for the these models that explains 
the origin of the $ \mathbb{Z}_2 $ symmetry breaking term and the effective 
Higgs-Yukawa terms, and (ii) extending this kind of models such that 
they also include the mass hierarchy of the other SM fermions.

\section{Composite two-Higgs doublet models With $ \mathbb{Z}_2 $ Symmetry}

\begin{table}[tb]
\begin{center}
{\renewcommand{\arraystretch}{1.2}
\begin{tabular}{c|c|c|c|c|c}
\hline\hline
    & $\mbox{G}_{\rm TC} $ & $\mbox{SU(3)}_{\rm C} $  & $\mbox{SU(2)}_{\rm L} $ & $\mbox{U(1)}_{\rm Y} $ & $ \mathbb{Z}_2 $ \\
    \hline\hline
 $ \Psi_1\equiv (\psi_1,\psi_2)^T $  & $\Box$ & $ \textbf{1}$&  $\Box$ &  $0$ &$ +1 $  \\ 
  $ \psi_3 $  & $\Box$ & $\textbf{1}$&  $\textbf{1}$&  $-1/2$ &$ +1 $  \\ 
  $ \psi_4 $  & $\Box$ & $\textbf{1}$&  $\textbf{1}$&  $+1/2$ &$ +1 $  \\
 $ \Psi_2\equiv (\psi_5,\psi_6)^T $  & $\Box$ & $ \textbf{1}$&  $\Box$ &  $0$ &$ -1 $ \\
\hline\hline
 $ Q_{L,3} \equiv (t_L,b_L)^T $  & $\textbf{1}$ & $\Box$&  $\Box$ &  $+1/6$ &$ +1 $  \\ 
  $ t_R $  & $\textbf{1}$ & $\Box$&  $\textbf{1}$&  $+2/3$ &$ +1 $  \\ 
  $ b_R $  & $\textbf{1}$ & $\Box$&  $\textbf{1}$&  $-1/3$ &$ -1 $ \\
\hline\hline
\end{tabular} }
\end{center}
\caption{The techni-fermions and the quarks in the SU(6)/Sp(6) template model labelled with their representations of $ \rm G_{\rm TC} \times SU(3)_C \times SU(2)_L \times U(1)_Y $ and parity under the $ {\mathbb{Z}_2} $ symmetry. } \label{tab:fermionssu6sp6}
\end{table}

We presently focus on CH models with  misalignment based on an underlying gauge description of strongly interacting fermions (techni-fermions). Different chiral symmetry breaking patterns in these CH models are discussed in Refs.~\cite{Witten:1983tx,Kosower:1984aw},
and we note the following minimal cosets with a Higgs candidate and custodial symmetry: SU(4)/Sp(4)~\cite{Galloway:2010bp}, SU(5)/SO(5)~\cite{Dugan:1984hq}, SU(6)/Sp(6)~\cite{Cai:2018tet}, SU(6)/SO(6)~\cite{Cacciapaglia:2019ixa}, and SU(4)$\times$SU(4)/SU(4)~\cite{Ma:2015gra}. Two composite Higgs doublets and a $\mathbb{Z}_2$ symmetry are present in the three latter cases~\cite{Cai:2018tet,Cacciapaglia:2019ixa,Ma:2017vzm}, where the coset $\SU(6)/\SP(6)$ generates the minimal number of pNGBs that simultaneously fulfils our requirements. With an unbroken  $\mathbb{Z}_2$ symmetry, this kind of model may also provide (a)symmetric dark matter candidates~\cite{Cai:2019cow}. Note that our proposal is rather general because the above requirements can also be fulfilled in other realisations that do not have a simple gauge-fermion underlying description, e.g. the models in Ref.~\cite{Mrazek:2011iu,Bertuzzo:2012ya}, and they can also be fulfilled in fundamental realisations.

We remind the reader that
the basic idea of natural composite Higgs models, based on 
vacuum misalignment~\cite{Contino:2010rs,Panico:2015jxa}, is that
the SM Higgs doublet will be identified as composite pNGBs.
These appear in the cosets $\mathrm{G}/\mathrm{H}$ of global symmetry groups. 
The model consists of $ N_f $ new Weyl techni-fermions 
which form a representation of a new strongly interacting ``technicolor'' 
gauge group $ G_{\rm TC} $. The choice of either real, pseudo-real or a complex representation, 
determines the breaking pattern of a global symmetry $\mathrm{G}$ to $\mathrm{H}$. The model 
is constructed such that the EW gauge symmetry $\SU(2)_L \times \UU(1)_Y$ 
is contained in $\mathrm{H}$ for a certain alignment, 
known as the EW unbroken alignment limit. 

This particular alignment, however, is not stable 
because there exists an explicit breaking of $\mathrm{G}$ in 
the form of gauge interactions, top couplings to the strong sector, 
and explicit masses for the techni-fermions, etc. The EW gauge interactions, 
the SM fermion couplings to the strong sector and the explicit masses of the 
techni-fermions contribute to the effective Higgs potential. These terms 
are responsible for generating a VEV of the Higgs and corresponds to a 
misalignment of the vacuum. We describe this 
by an angle, $\sin \theta =  v_{\rm EW}/(2\sqrt{2}f)$~\cite{Kaplan:1983fs}, 
where $ v_{\rm EW} =246$~GeV and $f$ is the decay constant of the pNGBs 
depending on the confinement of the underlying strong dynamics. From 
electroweak precision measurements~\cite{Agashe:2006at,Grojean:2013qca} 
this angle can be fixed typically to $\sin \theta \lesssim 0.2$, which also 
fixes $2\sqrt{2}f \gtrsim 1.2$~TeV, however, it may also be allowed for 
smaller scales~\cite{Ghosh:2015wiz,BuarqueFranzosi:2018eaj}.

If we assume that the VEV hierarchy of the Higgs doublets only arises from vacuum misalignment with two separate angles, $ \sin\theta_t = v_{\rm EW}/(2\sqrt{2}f) $ and $ \sin\theta_b = v_b/(2\sqrt{2}f) $, then we require a very fine-tuned angle
$ \theta_b \ll \theta_t \lesssim 0.2 $. We assume therefore the vacuum is only misaligned along the SM Higgs direction while a smaller tadpole VEV of the second doublet is provided by breaking the $\mathbb{Z}_2$ symmetry. The coset structure can be schematically represented by a $N_f \times N_f$ matrix,
\beq
\left( \begin{array}{c|c}
\mathrm{G}_0/\mathrm{H}_0 & \begin{array}{c} \mathbb{Z}_2\mbox{--odd} \\ \mbox{pNGBs} \end{array} \\ \hline
\begin{array}{c} \mathbb{Z}_2\mbox{--odd} \\ \mbox{pNGBs} \end{array} & \begin{array}{c} \mathbb{Z}_2\mbox{--even} \\ \mbox{pNGBs} \end{array}
\end{array} \right), \label{eq. Nf Nf matrix}
\eeq
where $\mathrm{G}_0/\mathrm{H}_0$ is one of the two minimal cosets 
$\SU(4)/\SP(4)$ or $\SU(5)/\SO(5)$, with one composite Higgs doublet. The $ \mathbb{Z}_2 $ symmetry can be understood in terms of the underlying techni-fermions 
$\psi_i$, $i = 1, \dots N_f$, that condense: $\psi_{5, \dots N_f}$ are $\mathbb{Z}_2\mbox{--odd}  $ while the techni-fermions that participate to the minimal coset are $\mathbb{Z}_2\mbox{--even}  $. Among the $\mathbb{Z}_2\mbox{--odd}  $ pNGBs must be contained the $\mathbb{Z}_2\mbox{--odd}  $ Higgs doublet $ H_b $.

\section{A Concrete Composite 2HDM}

In the following, we focus on the concrete $\SU(6)/\SP(6)$ model~\cite{Cai:2018tet} as a template for this mechanism. We assume four Weyl fermions are arranged in $\SU(2)_L$ doublets, $\Psi_1 \equiv (\psi_{1},\psi_{2})^T$ and $\Psi_2 \equiv (\psi_{5},\psi_{6})^T$, and two in $\SU(2)_L$ singlets, $\psi_{3,4}$, with hypercharges $ \mp 1/2 $. We have listed in Table~\ref{tab:fermionssu6sp6} the representations of the gauge groups and parity of the fermions of this model.

By assuming the Weyl fermions are in fundamental representation of the new strongly interacting gauge group $ \rm G_{\rm TC} = \rm SU(2)_{TC} $ or $ \rm Sp(N)_{\rm TC} $, which is the pseudo-real representation, we can then construct an $ \SU(6) $ flavour multiplet by arranging the six Weyl fermions into an $ \SU(6) $ vector $ \Psi \equiv (\psi^1,\psi^2,\psi^3,\psi^4,\psi^5,\psi^6)^T $. This results in the chiral symmetry breaking $ \SU(6)\rightarrow \text{Sp}(6) $ when the fermions confine. The fermions develop a non-trivial and antisymmetric vacuum condnesate~\cite{Galloway:2010bp}
\beq
\langle \Psi^I_{\alpha,a}\Psi^J_{\beta,b}\rangle\epsilon^{\alpha\beta}\epsilon^{ab}\sim \Phi^{IJ}_{\rm CH}.
\eeq 
$ \alpha,\beta $ are spinor indices, $ a,b $ are TC indices, and $ I,J $ are flavour indices. We will suppress the contractions of these indices for simplicity. The CH vacuum of the model, giving rise to the EW VEV of $ H_t^0 $ by misalignment, can be written as~\cite{Galloway:2010bp} \beq \label{EW vacuum matrix}
\Phi_{\text{CH}}=\begin{pmatrix}i\sigma_2 c_\theta &\mathbb{1}_2 s_\theta&0\\ -\mathbb{1}_2 s_\theta &-i\sigma_2 c_\theta&0\\0&0&i\sigma_2\end{pmatrix},
\eeq where from now on we use the definitions $ s_x \equiv \sin x $, $ c_x\equiv \cos x $ and $ t_x \equiv \tan x $.

\begin{table}[tb]
\begin{center}
{\renewcommand{\arraystretch}{1.2}
\begin{tabular}{c|c|c}
\hline\hline
    & $\begin{array}{c} \mbox{EW Vacuum} \\ (\theta = 0) \end{array}$ & $\begin{array}{c} \mbox{TC Vacuum} \\ (\theta = \pi/2) \end{array}$ \\
\hline\hline
$\displaystyle \mathrm{G}_0/\mathrm{H}_0$ &  $\begin{array}{c} H_t = (2,1/2)_+ \\ \eta = (1,0)_+ \end{array}$ & $\begin{array}{c} \phi = (h+i \eta)/\sqrt{2} \\ \pi^\pm\,, \;\; \pi^0 \end{array}$ \\
\hline
$\begin{array}{c} \mathbb{Z}_2\mbox{--odd} \\ \mbox{pNGBs} \end{array}$ & $\begin{array}{c} H_b = (2,1/2)_- \\ \Delta = (3,0)_- \\  \varphi^0 = (1,0)_- \end{array}$ &  $ \begin{array}{c}  \Xi_{1} = - H_b^0 + (\Delta^0 + i \varphi^0)/\sqrt{2} \\ \Xi_{2} =  (H_b^0)^\ast + (\Delta^0 - i \varphi^0)/\sqrt{2} \\ \Xi^\pm_{1} = \Delta^\pm - H_b^\pm \\ \Xi^\pm_{2} = \Delta^\pm + H_b^\pm  \end{array}$ \\
\hline
$\begin{array}{c} \mathbb{Z}_2\mbox{--even} \\ \mbox{pNGBs} \end{array}$ & $\eta' = (1,0)_+ $ & $\eta'$ \\ 
\hline\hline
\end{tabular} }\end{center}
\caption{The pNGBs in the template SU(6)/Sp(6) model in the EW unbroken labelled with their $ (\rm SU(2)_L, U(1)_Y)_{\mathbb{Z}_2}$ quantum numbers and in the TC vacuum. Note that $H_t = (\pi^+, (h+\ii \pi^0)/\sqrt{2})^T$ and $\pi^\pm, \pi^0$ are the Goldstones eaten by $W$ and $Z$.} \label{tab:su6sp6}
\end{table}

The chiral symmetry breaking $ \SU(6)\rightarrow \text{Sp}(6) $ results in 14 pNGBs, $ \pi_a $ with $ a=1,...,14 $, and thus 14 $ \SU(6) $ broken generators, $ X_a $. The Goldstone bosons around the EW vacuum are parametrized as $\Sigma=\exp[i\pi_aX_a/f]\Phi_{\text{CH}}$ with the decay constant $ f $ of them. This model preserves a $ \mathbb{Z}_2 $ symmetry generated by the $ \SU(6) $ matrix, which is 
\beq 
P = \text{Diag}(1,1,1,1,-1,-1), \label{eq: parity symmetry}
\eeq 
where the $ \mathbb{Z}_2\mbox{--odd} $ fields of the model are \beq 
H_b^0,\quad (H_b^0)^*, \quad H_b^{\pm}, \quad \Delta^0, \quad \Delta^\pm, \quad  \varphi^0, \quad b_R.
\eeq 
We have listed in Table~\ref{tab:su6sp6} the quantum numbers and parity of the pNGBs divided into the various groupings in Eq.~(\ref{eq. Nf Nf matrix}) for the EW unbroken ($ \theta = 0 $) and the TC vacuum ($ \theta = \pi/2 $). The neutral components of $H_{t,b}$ in unitary gauge are as follows:
\beq
H_t^{0}\equiv\frac{h}{\sqrt{2}},\quad H_b^{0}\equiv\frac{h_b+\ii\chi_b}{\sqrt{2}}. 
\eeq
 
In terms of the six-Weyl spinors, $\Psi $, under the global flavour group $\SU(6)$ the underlying fermionic Lagrangian  can be written as 
\beq
\mathcal{L}= \Psi^\dagger i \gamma^\mu D_\mu \Psi  
	 + \delta \mathcal{L}+\delta \mathcal{L}_m\,, 
\label{Basic Lagrangian (UV)}
\eeq
where the covariant derivatives involve the techni-gluons and the 
$\SU(2)_{L}$ and $ \UU(1)_Y $ gauge generators. The terms $ \delta \mathcal{L}$ are additional interactions including 
the four-fermion operators responsible for Yukawa couplings of the top and bottom and the $ \mathbb{Z}_2 $ symmetry breaking term given in Eqs.~(\ref{four fermion top and bottom}) and~(\ref{Z2 broken four fermion operator}).
Finally, we have collected the Lagrangian terms with vector-like masses
$ m_{1,2,3} $ for $\Psi$ in $ \delta \mathcal{L}_m$ and will consider 
       \beq
	\label{eq:MQMLambda}
 \delta \mathcal{L}_m= \frac{1}{2}\Psi^T M_\Psi \Psi +{\rm h.c.}, 
    \eeq 
    where $M_\Psi=\text{Diag}(im_1\sigma_2,-im_2\sigma_2,im_3\sigma_2)$. 

Below the condensation scale ($ \Lambda_{\rm TC}\sim 4\pi f $), Eq.~\eqref{Basic Lagrangian (UV)} yields the effective Lagrangian:
\beq
    \label{eq:effLag}
    \mathcal{L}_\mathrm{eff}=\mathcal{L}_{\mathrm{kin}}-V_{\mathrm{eff}}. 
\eeq The gauge-kinetic terms in $ \mathcal{L}_{\mathrm{kin}} $,  besides providing the kinetic terms for the pNGBs, induce the masses of the EW gauge bosons and their couplings with the pNGBs (including the SM Higgs), \beq \label{WZ masses and SM VEV}
&&m_W^2=2g_W^2f^2s_\theta^2,\quad m_Z^2=m_W^2/c^2_{\theta_W}, \\
&&g_{\rm hWW}=\sqrt{2}g^2_Wfs_\theta c_\theta=g_{  hWW}^{\rm SM}c_\theta,\quad g_{\rm hZZ}=g_{ hWW}/c^2_{\theta_W}, \nonumber
\eeq 
where $ v_{\rm EW}\equiv 2\sqrt{2}fs_\theta = 246~\text{GeV} $, $ g_W $ is the weak gauge 
coupling, and $ \theta_W $ is the Weinberg angle. 
The vacuum misalignment angle $ \theta $ parametrizes the 
corrections to the Higgs couplings to the EW gauge bosons and 
is constrained by LHC data \cite{deBlas:2018tjm}. 
This would require a small $ \theta $ ($ s_\theta \lesssim 0.3 $), 
but even smaller according to the electroweak precision 
measurements~\cite{Agashe:2006at,Grojean:2013qca} ($s_\theta \lesssim 0.2$). 
Misalignment of $ \theta $ to a small value is controlled by the 
contributions from the EW gauge interactions, the SM fermion couplings 
to the strong sector and the vector-like masses of the techni-fermions. 
These terms contribute all to the effective potential in Eq.~(\ref{eq:effLag}), 
\beq
V_{\text{eff}}&=&V_{\text{gauge}}+V_{\text{top}}+V_{\text{bottom}}+V_\text{m} +\dotsc \label{Potential 1}
\eeq

We further require operators that generate the top and 
bottom Yukawa couplings and the fermion loop contributions to
the potential in Eq.~(\ref{Potential 1}). 
These operators could be analogous to the four-fermion interactions
in \cite{Cacciapaglia:2014uja},
\beq \label{four fermion top and bottom}
\frac{y_{t}}{\Lambda_t^2}(Q_Lt_R^c)_\alpha^\dag(\Psi^TP_{t}^\alpha\Psi)+\frac{y_{b}}{\Lambda_b^2}(Q_Lb_R^c)_\alpha^\dag(\Psi^TP_{b}^\alpha\Psi),
\eeq 
where it is assumed $ y_f \equiv y_t = y_b $ and that there is one 
scale $ \Lambda_f \equiv \Lambda_t = \Lambda_b $ from an 
underlying mechanism which we for now leave unspecified. 
The spurions, $ P_{t,b}^\alpha $, project out the EW components 
such that $ \Psi^TP_{t,b}^\alpha\Psi $ transform as the two Higgs doublets. 
When the techni-fermions condense, these terms generate the top contribution 
to the effective potential in Eq.~(\ref{Potential 1}) and the following operators: 
\beq \label{condense Yukawa operators}
g f\left[ (Q_Lt_R^c)_\alpha^\dag \text{Tr}[P_{t}^\alpha \Sigma]+ (Q_Lb_R^c)_\alpha^\dag \text{Tr}[P_{b}^\alpha \Sigma]\right]
\eeq 
with 
$ g \equiv  4\pi N A (\Lambda_{\text{TC}}/\Lambda_f)^2 y_f $~\cite{Hill:2002ap} 
where $ A $ is an integration constant arising from the condensation 
and $ N $ is the number of technicolors. 
So far, the model only generates the top quark mass,
while the bottom quark is massless,
\beq \label{mt mb masses and couplings}
&&m_t=gfs_\theta=\frac{g v_{\rm EW}}{2\sqrt{2}},\quad \quad  m_b = 0. 
\eeq 

We generate a tadpole VEV for $ h_b $ by adding 
an operator that weakly mixes $ H_t $ and $ H_b $ and thus 
softly breaking the $ \mathbb{Z}_2 $ symmetry in Eq.~(\ref{eq: parity symmetry}).  
We introduce a four-fermion interaction 
that generates the $ \mathbb{Z}_2 $ breaking terms, 
\beq
\frac{g_Z}{\Lambda_Z^2} (\Psi^T S_{15}\Psi)(\Psi^T S_{5}\Psi), \label{Z2 broken four fermion operator}
\eeq where $ S_{5} $ and $ S_{15} $ are unbroken generators of the unbroken global group Sp(6), which are \beq\label{unbroken}
		S_{5}=\frac{1}{2}\begin{pmatrix}0&0&0\\0&-\sigma_{2}^T&0\\0&0&0\end{pmatrix}, \quad S_{15}=\frac{1}{2\sqrt{2}}\begin{pmatrix}0&0&\sigma_{2}\\0&0&0\\\sigma^{2}&0&0\end{pmatrix}.\nonumber
\eeq 
After the condensation, we obtain the following $ \mathbb{Z}_2 $ breaking terms 
in Eq.~(\ref{Potential 1}), \beq \label{condense Z2 breaking terms}
V_{\text{Z}}&=&  \tilde{g}_Z f^4 \text{Tr}[S_{15}\Sigma^\dagger S_5 \Sigma]\\
&=&-\frac{\tilde{g}_Z  f^3}{8 s_{\theta/2}}s_\theta^2 h_b - \frac{\tilde{g}_Z f^2}{32\sqrt{2}}(c_{\theta/2}+3c_{3\theta/2})hh_b +\dotsc \nonumber
\eeq with $ \tilde{g}_Z\equiv (4\pi)^4 N^2 A^2  (\Lambda_{\text{TC}}/\Lambda_\text{Z})^2 g_Z$~\cite{Hill:2002ap}. 
The above operator provides a mixing term between $ h $ and $ h_b $, 
and therefore a tadpole VEV for $ h_b $. 
By minimizing the total Higgs potential in Eqs.~(\ref{Potential 1}) and~(\ref{condense Z2 breaking terms}), we obtain the tadpole VEV of $ h_b $, which is determined by \beq \label{vb}
&& v_b = \frac{g \tilde{g}_Z t_{\theta/2}^{-1} f^2 s_\theta^2}{16\sqrt{2}\pi Zm_{23}+2\sqrt{2}f(C_g g^2_W(3+t_{\theta_W}^2)-2C_f g^2)c_\theta}, \nonumber
\eeq where $ m_{ij}\equiv m_i + m_j $ with $ i,j =1,2,3 $. 
The loop factors $ C_{g,f} $ and the constant $ Z $ occurring in the gauge, 
SM fermion loop corrections and effective vector-like mass terms in 
the effective potential (Eq.~\ref{Potential 1}), respectively, are 
non-perturbative $ \mathcal{O}(1) $ coefficients. These coefficients can 
be suggested by lattice simulations of the underlying strong dynamics. 

Hence, both the top and bottom quarks obtain masses, 
\beq \label{final top bottom mass}
&&m_t=\frac{g v_{\rm EW}}{2\sqrt{2}},\quad \quad  m_b = \frac{g v_b}{2\sqrt{2}}c_{\theta/2}.
\eeq  
Notice
that the tadpole VEV $ v_b $ can be reduced 
by increasing the vector-like mass  $ m_3 $ for the second 
techni-fermion doublet $ \Psi_2 $, and the bottom 
mass can thus be tuned down to its observed value.

\section{Numerical Results of the Concrete Composite 2HDM}

We turn to a numerical analysis of the mass spectrum of the concrete composite 2HDM that generates the top-bottom mass hierarchy. For simplicity, we assume the vector-like masses in the minimal coset are equal, $ m_1 = m_2 $, and the unknown constant $ Z = 1 $ (expected to be $ \mathcal{O}(1) $). The expressions for the EW VEV (Eq.~(\ref{WZ masses and SM VEV})), the SM Higgs mass, and the top and bottom mass (Eq.~(\ref{final top bottom mass})) can be fixed to their observed values~\cite{Tanabashi:2018oca}, and $ m_{12} $ can be eliminated by the vacuum misalignment condition for $ \theta $ from the minimization of the effective potential in Eq.~(\ref{Potential 1}). Here, we have collected these expressions: \beq
&&v_{\rm EW}=2\sqrt{2}f s_\theta \approx 246~\mathrm{GeV}, \quad \quad m_{\tilde{h}} \approx 125~\mathrm{GeV}, \nonumber \\ && m_t = \frac{g v_{\rm EW}}{2\sqrt{2}}\approx 173~\mathrm{GeV},\quad \quad m_b = \frac{g v_b}{2\sqrt{2}}c_{\theta/2} \approx 4.2~\mathrm{GeV}, 
\nonumber \\ && m_{12} = \frac{f(2C_f g^2 -C_g g_W^2(3+t_{\theta_W}^2))c_\theta }{8\pi Z}, \nonumber
\eeq where $ m_{\tilde{h}} $ is the mass of the physical SM Higgs, $ \tilde{h} $, which is the mass eigenstate of a mass matrix $ M_0^2 $ in the basis $ (h,h_b,\Delta^0) $. The SM Higgs $ \tilde{h} $ consists mostly of the $ \mathbb{Z}_2 $-even field $ h $ and less of the $ \mathbb{Z}_2 $-odd field $ h_b $ while the $ \mathbb{Z}_2 $-even field $ h_b $ also mixes with the neutral field of the triplet, $ \Delta^0 $. For further simplicity, we also fix the vacuum misalignment angle to $ s_\theta =0.1 $ ($ \Lambda_{\rm TC} \sim 4\pi f=10.9~\mathrm{TeV}$) such that we only have the coupling of the $ \mathbb{Z}_2 $ breaking terms, $ \tilde{g}_Z $, as free parameter.

\begin{figure}[t!]
	\centering
	\includegraphics[width=0.47\textwidth]{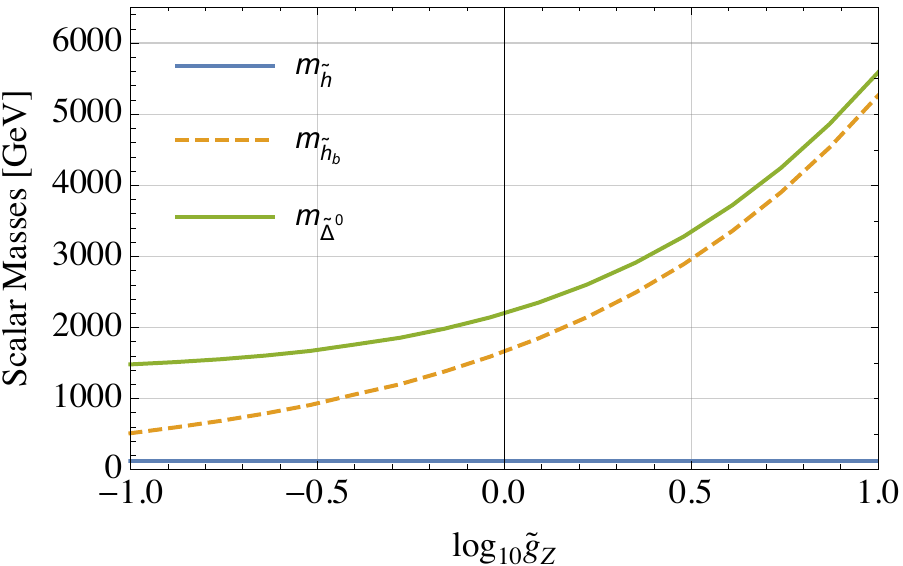}
	\phantom{o}\includegraphics[width=0.455\textwidth]{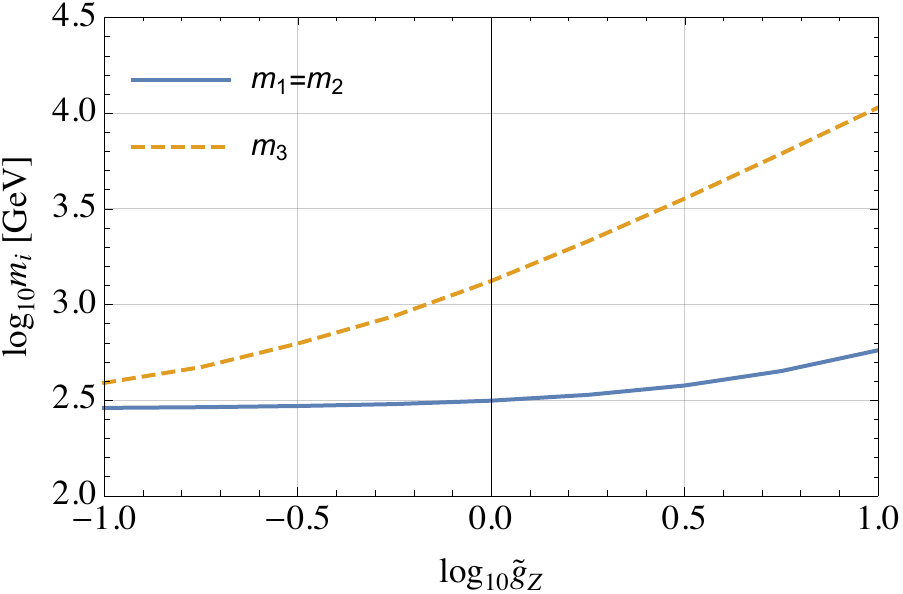}
	\caption{\textbf{Upper panel:} The masses of the Higgs bosons ($ \tilde{h} $ and $ \tilde{h}_b $) and the mostly neutral charged component of the triplet ($ \tilde{\Delta}^0 $) are shown for varying coupling of the $ \mathbb{Z}_2 $ breaking terms, $ \tilde{g}_Z $. \textbf{Lower panel:} The vector-like masses of the techni-fermions, $ m_i $ with $ i=1,2,3 $, for varying $ \tilde{g}_Z $. }
	\label{neutral masses}
\end{figure}

In Figure~\ref{neutral masses}, the mass spectrum of the neutral 
fields (upper panel) and the vector-like masses of the techni-fermions (lower panel) are shown for varying $ \tilde{g}_Z $ for fixed $ m_{\tilde{h}}=125 $ GeV. 
The ratio of the non-perturbative coefficients for the fermion loops and 
the gauge loops is $ C_{f}/C_{g} = 2...4 $ for the interval $ \tilde{g}_Z = 0.1...10 $ and therefore they can be $ \mathcal{O}(1) $. 

For $ \tilde{g}_Z = 1 $, the mass splitting between the vector-like masses are $ m_1=m_2 = 314~\mathrm{GeV} $ and $ m_3 = 1326~\mathrm{GeV} $. This mass splitting generates the observed masses of the top and bottom quark. 
In this case, the masses of all the composite fields are above 1 TeV (except of the SM Higgs), while the mass of $ \tilde{h}_b $ is $ 1.7 $ TeV. In Refs.~\cite{Hill:2019ldq,Hill:2019cce}, it is emphasized that the LHC may already have the capability of ruling out an $ h_b $ with $ g\sim 1 $ of mass $ \sim 1 $ TeV with current integrated luminosities, $ \sim 200 \rm ~fb^{-1} $, set by the constraints of the $ h_b $ production with 
same couplings like in our theory. Our theory is therefore testable and should provide motivation to go further and deeper into the energy 
frontier with LHC upgrades with a 100 TeV $ pp $ and/or high energy lepton colliders. 

Finally, we estimate  the couplings $ y_f $ and $ g_Z $ 
for the four-fermion operators in Eqs.~(\ref{four fermion top and bottom}) 
and~(\ref{Z2 broken four fermion operator}). For  $ A\sim 1 $ and $ N=2 $, we find that 
\beq
&&y_f\sim 0.1 \left(\frac{\Lambda_{f}}{\Lambda_{\rm TC}}\right)^2, \quad g_Z\sim 10^{-5} \left(\frac{\Lambda_{Z}}{\Lambda_{\rm TC}}\right)^2 \tilde{g}_Z. 
\eeq 
Assuming $ y_f \sim 1 $, we obtain the scale $ \Lambda_f \sim 40 $ TeV 
for $ s_\theta = 0.1 $. Then for $ \tilde{g}_Z=1 $ and $ \Lambda_Z= \Lambda_f \sim 40 $ TeV,
we estimate $ g_Z \sim  10^{-4} $. 
The smallness of this coupling is 
technically natural according to 't Hooft's naturalness principle \cite{tHooft:1979rat} 
because the $ \mathbb{Z}_2 $ symmetry of the 
condensate is recovered when we take the limit $ g_Z\rightarrow 0 $.
Note, however, small technically natural couplings such as $g_Z $ arise dynamically
from power-law suppressions of large initial couplings such as  $ \tilde{g}_Z $.


\section{Conclusions and Future Work}

We have relegated the detailed study of the origin of the four-fermion operators in Eqs.~(\ref{four fermion top and bottom}) and (\ref{Z2 broken four fermion operator}) to future work. Probably these operators can be derived from a 
Fundamental Partial Compositeness (FPC) theory like in Refs.~\cite{Sannino:2016sfx,Cacciapaglia:2017cdi} with global symmetry group $ \SU(6)_{\mathcal{F}}\times \SP(12)_{\mathcal{S}} $. In constrast to Ref.~\cite{Cacciapaglia:2017cdi}, there will be two complex color- and TC-charged scalars, a $ \mathbb{Z}_2 $-even $ S_t $ and -odd $ S_b $ that are in the fundamental representation of a global symmetry $ \SP(12)_{\mathcal{S}} $ while the techni-fermions transform under $ \SU(6)_{\mathcal{F}} $. We can potentially introduce new renormalizable terms with one coupling constant involving the techni-fermions, techni-scalars and the SM fermion. These terms will dynamically generate similar Higgs-Yukawa terms like in Eq.~(\ref{condense Yukawa operators}). By introducing a weak $ \mathbb{Z}_2 $-breaking term that mixes the fundamental fields $ S_t $ and $ S_b $ will potentially generate effective operators similar to them generated by the four-vector operator in Eq.~(\ref{Z2 broken four fermion operator}).

The concrete $ \SU(6)/\SP(6) $ model can  be extended to describe the masses
of the other SM fermions by adding a $ \SU(2)_L $ techni-fermion doublet 
for each SM fermion. Such a CH model will contain a $ \SU(2)_L $ techni-fermion doublet $ \Psi_i $ for each SM fermion with $ i=b,c,s,... $, one $ \SU(2)_L $ doublet $ \Psi_t $ for the top quark and two $ \SU(2)_L $ techni-fermion singlets. For each $ \Psi_i $, there will exist a $ \mathbb{Z}_2^{(i)} $ symmetry of the condensate. Assuming these $ \mathbb{Z}_2^{(i)} $ symmetries are softly broken by four-fermion operators with one small coupling $ g_Z $, we create a mass hierarchy between the SM fermion masses controlled by the hierarchy of the vector-like masses of the techni-fermion doublets.


In conclusion, we have presented a novel mechanism for generation of the top-bottom mass hierarchy with one natural Higgs-Yukawa coupling, $ g \simeq 2  $. This top-bottom mass hierarchy is provided by breaking a $ \mathbb{Z}_2 $ symmetry of the condensate of new confining techni-fermions and in principle controlled by the vector-like masses of these techni-fermions. This mechanism can also be present in a wide variety of other models based on vacuum misalignment. 

We have considered an $ \SU(6)/\SP(6) $ CH template model where we showed this model can naturally explain the top-bottom mass hierarchy without fine-tuning of Yukawa coupling constants. Finally, we have briefly discussed the possibility of an underlying theory generating the four-fermion operators in this template model, and furthermore the possibility to extend it to describe the hierarchy of the other SM fermion masses.

\section*{Acknowledgements}
This work was done at Fermilab, 
operated by Fermi Research Alliance, LLC under Contract No. DE-AC02-07CH11359 with the United States Department of Energy. MR would like to thank Fermilab and Caltech for hosting him during the writing of this paper. 
MR acknowledges partial funding from The Council For Independent Research, grant number DFF 6108-00623. The CP3-Origins centre is partially funded by the Danish National Research Foundation, grant number DNRF90. 

%
%

\end{document}